\documentclass[eqsecnum,aps]{revtex4}
\usepackage{amsmath,amssymb,amsthm,bm}
\usepackage{psfrag}
\usepackage[dvips]{graphicx}
\usepackage[utf8]{inputenc} 
\usepackage{hyperref}

\begin{document}
\title{ Fractional topological charges and the lowest Dirac modes}
\author{Seyed Mohsen Hosseini Nejad}
\email{smhosseininejad@ut.ac.ir}
\affiliation{
Faculty of Physics, Semnan University, P.O. Box 35131-19111, Semnan, Iran}

\begin{abstract}

 We introduce vortex configurations with fractional topological charges where one unicolor or colorful intersection of two perpendicular vortex pairs contributes to the topological charge of the configurations. Using both, the overlap and asqtad staggered fermion formulations, the lowest modes of the Dirac operator on the noninteger $Q$ configurations are studied in the fundamental and adjoint representations. We analyze the behavior of the fundamental and adjoint fermions in the background of the topological charge contributions of $|Q|=0.5$.
\end{abstract}

\pacs{11.15.Ha, 12.38.Aw, 12.38.Lg, 12.39.Pn}

\maketitle

\section{INTRODUCTION}\label{Sect0}
Understanding quark confinement and spontaneous chiral symmetry breaking (SCSB) and the dynamical mechanism behind them are the big challenges in nonperturbative QCD. Lattice simulations and infrared models have indicated that the random fluctuations of the number of center vortices which are quantized magnetic fluxes in terms of the nontrivial center elements linked to the Wilson loop leads to quark confinement~\cite{DelDebbio:1996mh,Langfeld:1997jx,DelDebbio:1997ke,Langfeld:1998cz,Kovacs:1998xm,Engelhardt:1999wr,Betti:2017zcm,Oxman:2012ej,Engelhardt:1999fd,Bertle:2002mm,Engelhardt:2003wm,Engelhardt:2010ft,Hollwieser:2010mj,Hollwieser:2011uj,Schweigler:2012ae,Hollwieser:2012kb,Hollwieser:2014lxa,Altarawneh:2015bya,Hollwieser:2015qea,Altarawneh:2016ped}. In addition, lattice
simulations have shown that center vortices are also responsible for topological charge and SCSB~\cite{Engelhardt:2000wc,Bertle:2001xd,Bruckmann:2003yd,hollwieser:2008tq,Hollwieser:2009wka,Hollwieser:2017xmn,Bowman:2010zr,Hollwieser:2013xja,Brambilla:2014jmp,Hoellwieser:2014isa,Trewartha:2014ona,Hollwieser:2014mxa,Trewartha:2015nna,Nejad:2015aia,Hollwieser:2015koa,Faber:2017alm,Nejad:2016fcl,Nejad:2018pfl}.
The vortex intersections could contribute to the topological charge density
\cite{Hollwieser:2011uj}. Moreover, the 
color structure of vortices could contribute to the topological charge density too \cite{Schweigler:2012ae,Nejad:2015aia}. We studied colorful vortex planes in Ref. \cite{Nejad:2015aia} where the links of the colorful region are distributed over the full
SU($2$) gauge group. In addition to center vortices, the condensation of the Abelian monopoles also leads to the quark confinement in the dual superconductor scenario \cite{tHooft:1981bkw,Mandelstam:1976tq,Nambu:1974zg,Baker:1991bc,Baker:1997bg,Baker:1998jw,Konishi:2018gqv,Oxman:2017boz}. In addition, although instantons which carry an integer topological charge of modulus one fail to produce confinement (see reviews in e.g.~\cite{Schafer:1996wv,Diakonov:2002fq}), the idea of calorons which are instantons at finite temperature is probably the most promising current version of monopole confinement in pure non-Abelian gauge theories. The monopole constituents (dyons) of the KvBLL calorons \cite{Kraan:1998pm,Kraan:1998sn,Lee:1998bb,Lopez-Ruiz:2016bjl} are a source of both electric and magnetic fields which have fractional topological charge $Q_{dyon}=1/N$ for the SU($N$) gauge theory. Also, merons as another monopole degree of freedom carrying fractional topological charge may confine quarks \cite{Callan:1978bm,Callan:1977gz,Callan:1977qs}. Taken together all this motivates analyzing the fractional topological charges for the understanding of SCSB. An integer topological charge attracts the fundamental zero modes and would-be zero modes from opposite topological charges via interactions contribute to near-zero modes. In this work, we investigate the influence of fractional topological charges from center vortices on fermions.  

We study vortex fields with fractional topological charge which are
a combination of two antiparallel plane vortex pairs intersecting at four points but only one of these points contributes to the topological charge. The intersection points are considered unicolor or colorful intersections. The links of the unicolor intersection are distributed over a U($1$)-subgroup and the colorful intersection is constructed through locating the colorful region around the intersection point. 

These special configurations provide an opportunity to study the effect of the fractional topological charges on fermions. We analyze the lowest fundamental and adjoint modes of the overlap and asqtad staggered Dirac operator \cite{Narayanan:1993ss,Narayanan:1994gw,Neuberger:1997fp,Edwards:1998yw} for the configurations with fractional topological charge. According to the Atiyah-Singer index theorem
\cite{Atiyah:1971rm,Brown:1977bj,Adams:2000rn}, the difference between the number of left- and right-handed overlap fundamental zero modes in the background of a gauge field with integer topological charge $Q\neq 0$ is $|Q|$ and the one of the adjoint representation is $2N|Q|$ where $N= 2$ for SU($2$) gauge group.  In addition, the difference between the number of left- and right-handed (asqtad) staggered zero modes in the background of a gauge field with the integer topological charge $Q\neq 0$ is $|2Q|$ for the fundamental
and $|8Q|$ for the adjoint representation. We check the index theorem for fractional topological charges and study the lowest Dirac modes in the background of the noninteger $Q$ configurations. 

This paper is organized as follows. In Sec. \ref{Sect1}, vortex configurations with fractional topological charges are studied on the lattice. In Sec. \ref{Sect2}, we discuss the eigenmodes and eigenvalues of the Dirac operator in the fundamental and adjoint representations for the noninteger $Q$ configurations. In Sec.~\ref{Sect3}, we summarize the main points of our study.

\section{Vortex configurations with fractional topological charges}\label{Sect1}

We investigate plane vortices as classical configurations for SU($2$) lattice gauge theory \cite{Jordan:2007ff,Hollwieser:2011uj}. The plane vortices are parallel to two of the coordinate axes and occur in pairs of parallel sheets by using periodic boundary conditions (bcs) for the gauge fields. We use two different
arrangements of vortex sheets, $xy$- and $zt$-planes. The nontrivial links of unicolor plane vortices varying in a U($1$) subgroup of SU($2$) are
\begin{equation}
U_\mu=\exp(i \alpha \sigma_a),
\label{def_rn}
\end{equation}
where $\sigma_a$ ($a=x,y,z$) are the Pauli matrices. For $xy$-vortices, $t$-links in one $t$-slice $t_\perp$ and for $zt$-vortices, $y$-links in one $y$-slice $y_\perp$ are nontrivial.
The orientation of the plane vortices are determined by the gradient of the angle $\alpha$. Vortex pairs with the same vortex orientation are called parallel vortices and vortex pairs of opposite flux direction are called antiparallel. For $xy$-vortices, the angle $\alpha$ is chosen as a linear function of $z$, the coordinate perpendicular to the vortex, as the following \cite{Hollwieser:2011uj}
\begin{equation}\label{eq:phi-pl0}
\alpha(z)=\begin{cases}0&0<z\leq z_1-d,\\ 
                \frac{\pi}{2d}[z-(z_1-d)]&z_1-d< z\leq z_1+d,\\ 
                \pi&z_1+d<z\leq z_2-d,\\ 
                \pi\left[1-\frac{z-(z_2-d)}{2d}\right]&z_2-d<z\leq z_2+d,\\ 
                0&z_2+d<z\leq N_z.\end{cases}
\end{equation}
The parallel sheets of plane pair, which are opposite vortex orientations, have thickness of $2d$ around $z_1$ and $z_2$.  

For $zt$-vortices, the angle $\alpha$ is chosen the same as $xy$-vortices but a linear function of $x$.
The gluonic topological charge of these unicolor configurations is zero. The plane vortices could contribute to the topological charge density through the color structure and intersections. 

The colorful $xy$-plane vortices are introduced in Ref. \cite{Nejad:2015aia}. The color structure is considered for the first vortex sheet of the $xy$-plane vortices by the links \cite{Schweigler:2012ae,Nejad:2015aia}

\begin{equation}\label{originallinks2}
\quad U_{i}(x)=
\begin{cases}\left[g(\vec r+\hat i)\,g(\vec r)^\dagger\right]^{(t-1)/\Delta t}
& \mathrm{for}\quad1<t<1+\Delta t,\\
g(\vec{r}+\hat i)\,g\left(\vec r\right)^{\dagger}
& \mathrm{for}\quad1+\Delta t\leq t\leq t_g,\\
\mathbf 1&\mathrm{else},
\end{cases},~~~~~~U_4(x)= 
\begin{cases}
g(\vec r)^\dagger&\mathrm{for}\quad t=t_g,\\
 \mathbf 1&\mathrm{elsewhere}.\end{cases}
\end{equation}

$\Delta t$ is the duration of the transition between two vacua and
\begin{equation}\label{eq:sphv}
g(\vec{r})= 
\begin{cases}
\mathrm e^{\mathrm -i\alpha(z)\vec n\cdot\vec\sigma}\quad&\mathrm{for}\quad
z_1-d\leq z\leq z_1+d\quad\mathrm{and}\quad 0\leq\rho\leq R,
\\\mathrm e^{\mathrm -i \alpha(z)\sigma_3}  & \mathrm{else},\end{cases}
\end{equation}
where 
\begin{equation}\begin{aligned}\label{DefColVort}
&\vec n\cdot\vec\sigma=\sigma_x\,\sin\theta(\rho)\cos\phi(x,y)
+\sigma_y\,\sin\theta(\rho)\sin\phi(x,y)+\sigma_z\,\cos\theta(\rho),\\
&\rho=\sqrt{(x-x_0)^2+(y-y_0)^2},\quad
\theta(\rho)=\pi(1-\frac{\rho}{R})H(R-\rho)\;\in\,[0,\pi],\quad
\phi=\arctan_2\frac{y-y_0}{x-x_0}\;\in\,[0,2\pi).
\end{aligned}\end{equation}
$H$ is the Heaviside step function and $\vec{r}$ denotes  the spatial components. The colorful region which is a cylindrical region is located in the range $0 \leq \rho \leq R$ and $z_1-d\leq z \leq z_1+d$ with the center at $(x_0,y_0)$ and its lattice links are distributed over the full SU($2$) gauge group. For $\Delta t=1$, the colorful $xy$-vortices are not smoothed in temporal direction and the topological charge is zero which is a lattice artifact. Increasing the smoothing region $\Delta t$ of the colorful vortex, the topological
charge approaches $Q=-1$. The colorful $xy$-vortex is a fast vacuum to vacuum transition ($\Delta t=1$) which could be smoothed to show instanton like behavior~\cite{Schweigler:2012ae}.

Now, we study configurations with fractional topological charges. According to the topological charge definition:
\begin{gather}
  Q = - \frac{1}{32\pi^2} \int d^4x \, \epsilon_{\mu\nu\alpha\beta} \mbox{tr}[{\cal F}_{\alpha\beta} {\cal F}_{\mu\nu} ] = \frac{1}{4\pi^2} \int d^4x \, \vec E^a \cdot \vec B^a, \label{eq:qlatq}
\end{gather}
when a configuration in a region has both electric field ($\vec E^a$) and magnetic field ($\vec B^a$) with the same spatial and color ($\sigma_a$) directions, it contributes to the topological charge. On the lattice, ${\cal F}_{\mu\nu}$ is expressed in terms of the plaquette field
$P_{\mu\nu}= U_{\mu}(x)U_{\nu}(x +\mu)U^\dagger_{\mu}(x+\nu)U^\dagger_{\nu}(x)$. The $xy$-vortices with $U_t$ links, given in Eq.~(\ref{def_rn}), bear only
nontrivial $zt$-plaquettes ($P_{zt}$), {\it i.e.}, an electric field $E^a_z$, while
$zt$-vortices with $U_y$ links have nontrivial $xy$-plaquettes ($P_{xy}$) corresponding to a magnetic
field $B^a_z$. Crossing $E^i_z$-plaquettes ($P_{zt}$) of $xy$-vortices with color direction $\sigma_i$ and $B^i_z$-plaquettes ($P_{zt}$) of $zt$-vortices with color direction $\sigma_j$ ($j\neq i$) do not contribute to the topological charge.

\begin{figure}[h!] 
\centering
\includegraphics[width=0.48\columnwidth]{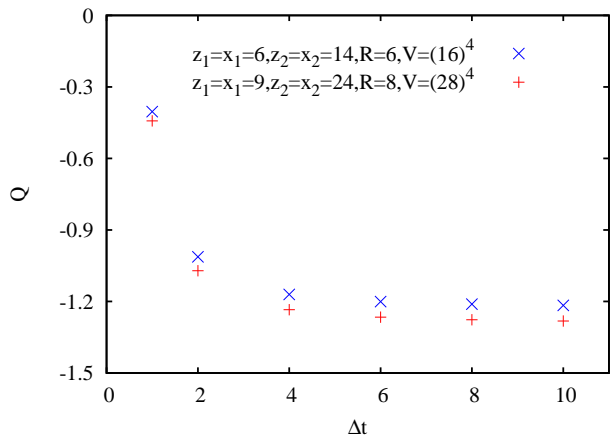}
\caption{The total topological charge $Q$ of the vortex configuration corresponding to Fig.~\ref{fig:2}(b). The gluonic topological
charge approaches $Q=-1.5$ by increasing the smoothing region $\Delta t$ of the colorful vortex and the radius $R$ of the colorful region through increasing the lattice size. }
\label{fig:1}
\end{figure}

\begin{figure}[h!] 
\centering
a)\includegraphics[width=0.4\columnwidth]{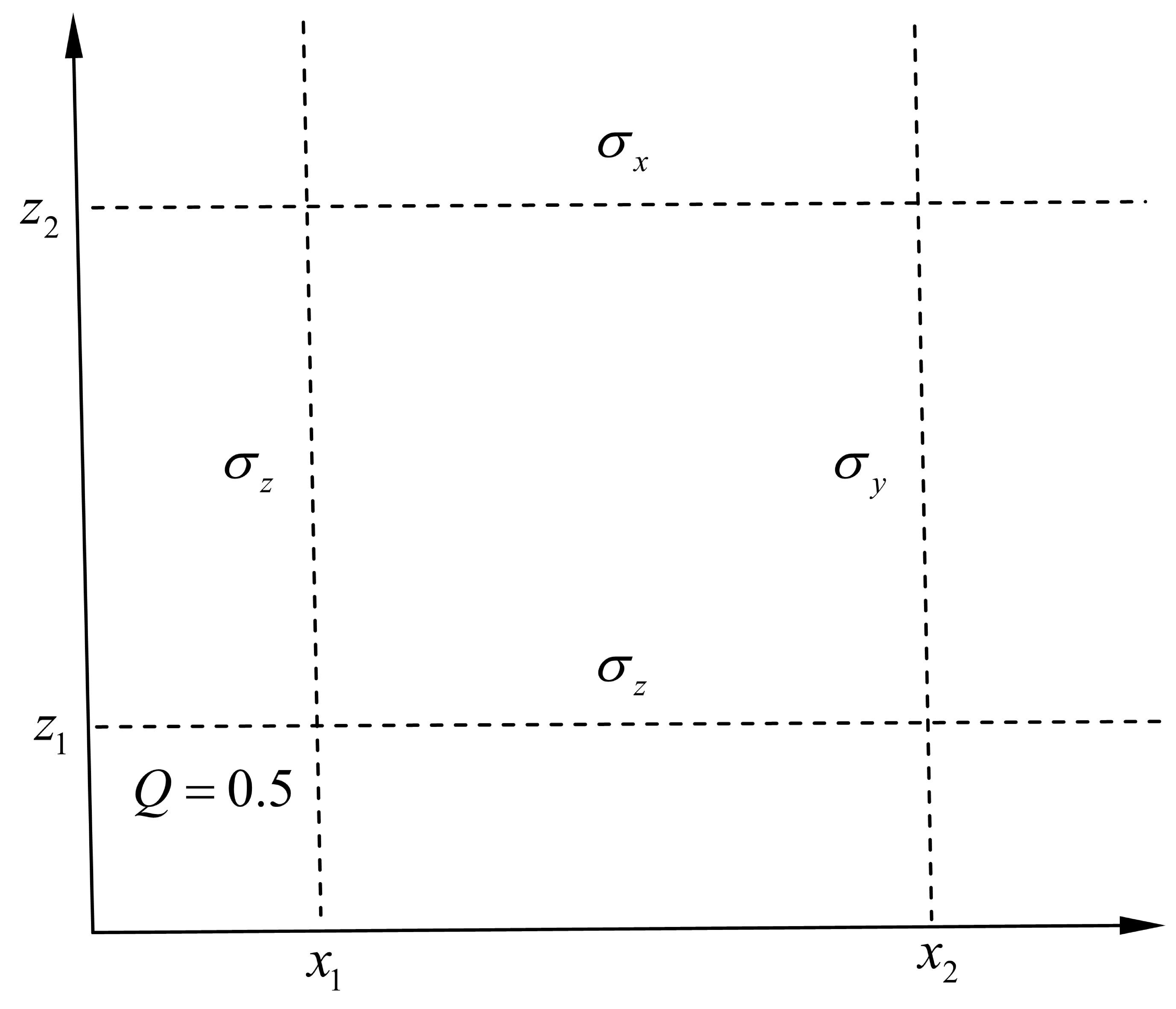}\hspace{1cm}
b)\includegraphics[width=0.4\columnwidth]{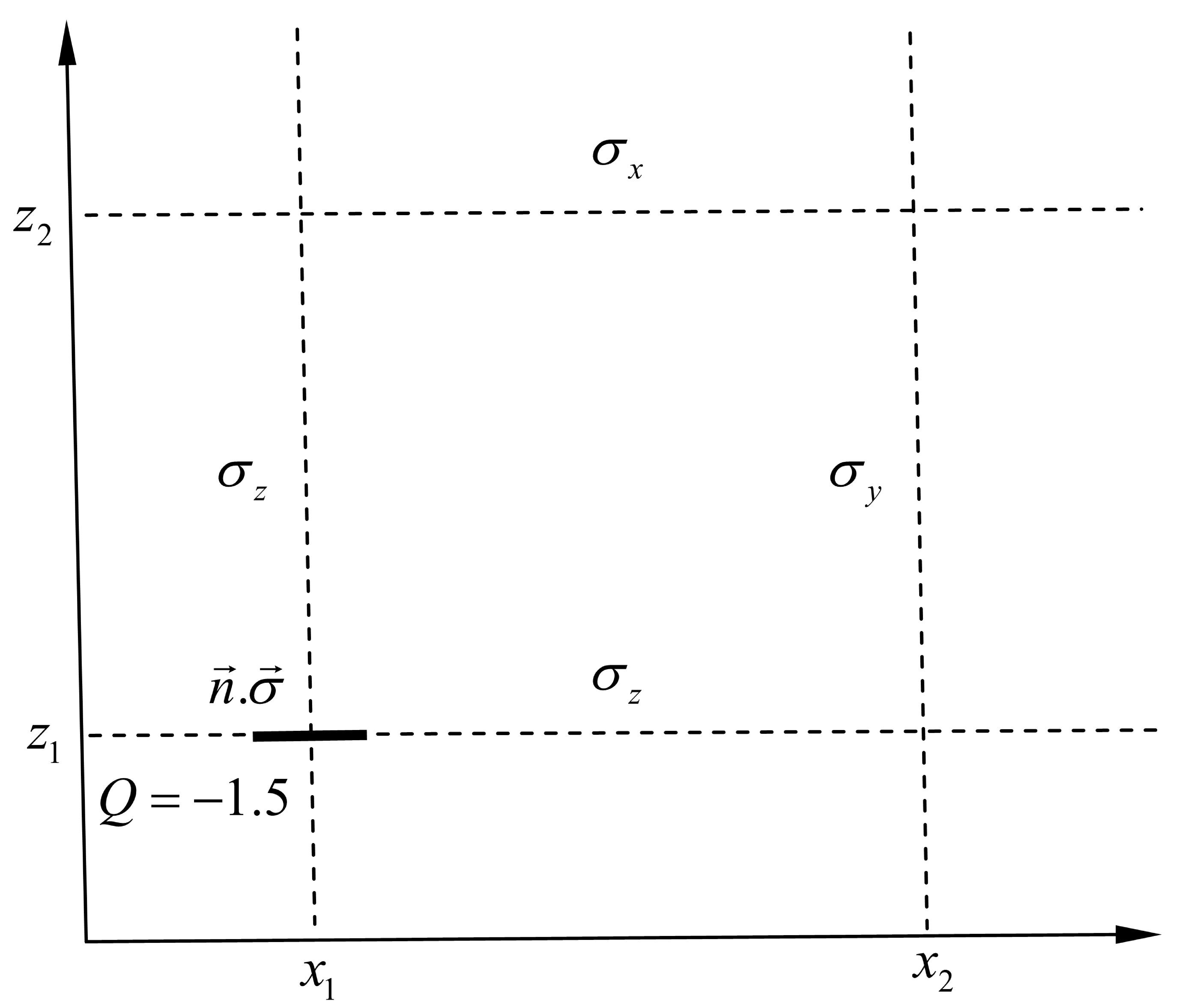}
\caption{The geometry and the topological charge contribution of the intersection points of two antiparallel vortex pairs. $\sigma_i$ on the vortices (dashed lines) means the links of these vortices are distributed over the $\sigma_i$-generated U($1$) subgroup of SU($2$). In diagram (b), the bold black line at the intersection point $(x_1,z_1)$ indicates that the unicolor vortex is substituted in this region by a colorful region. $\vec n\cdot\vec\sigma$ on the colorful region (bold black line) means the links belong to full SU($2$) gauge group. In both diagrams, the intersection point around $(x_1,z_1)$ contributes to the topological charge while the other intersection points have $Q=0$.}
\label{fig:2}
\end{figure}

Now we intersect two antiparallel vortex pairs of the $xy$- and $zt$-vortices with vortex centers $z_{1,2}$ and $x_{1,2}$ respectively. When the vortex sheets of an intersection point have the same color ($\sigma_a$) directions, the intersection point gives rise to a lump of topological charge $Q=\pm0.5$ \cite{Reinhardt:2002cm}. We choose the first (second) vortex sheet of the xy-vortices to rotate $\alpha$ in $\sigma_z$ ($\sigma_x$) and the first (second) vortex sheet of the zt-vortices to rotate $\alpha$ in $\sigma_z$ ($\sigma_y$). Therefore, intersection point around ($x_1,z_1$) carries the topological charge $Q=0.5$ while the other three intersection points have $Q=0$ and therefore sum up to a total topological charge $Q=0.5$. Now, for this configuration, the $Q=0.5$ configuration, the unicolor region of the first vortex for the $xy$-vortices around the point ($x_1,z_1$) is substituted by a colorful region. Figure~\ref{fig:1} shows the total topological charge of the configuration as a function of $\Delta t$ for two values of $R$ and increasing lattice sizes. As shown, the total topological charge converges to $Q=-1.5$ by increasing $\Delta t$ as well as increasing the radius $R$ of the colorful region, called the $Q=-1.5$ configuration. Schematic diagrams for the intersection planes of the $Q=0.5$ and $Q=-1.5$ configurations are plotted in Fig.~\ref{fig:2}.

Now, we study in detail the $Q=-1.5$ configuration. For $\Delta t=1$, the total topological charge of this configuration converges to $Q=-0.5$. 

In Ref. \cite{Nejad:2015aia}, we calculated the continuum action $S$ for the colorful cylindrical region of $xy$-vortices as
\begin{equation}
\label{action}
\frac{S(\Delta t)}{S_\mathrm{Inst}}=\frac{0.51\,\Delta t}{R}
+\frac{1.37\,R}{\Delta t}
\end{equation}
where $S$ is corresponding to the colorful region with thickness $d=R=7$ on a $28^3\times40$ lattice and the instanton action $S_\mathrm{Inst}=8\pi^2/g^2$. The minimum value of the action $S$ is reached around $R=\Delta t$ with 1.68~$S_\textrm{Inst}$. The first term in the action $S$, given in Eq.~(\ref{action}), represents the magnetic and the second term the electric contributions to the action. The action is purely electric for $\Delta t \to 0$. One gets the topological charge $Q=-1$ in the continuum for the sharp $xy$-vortices while the topological charge contribution for this configuration is zero on the lattice \cite{Nejad:2015aia}. Therefore, for the fast transition in temporal direction, both electric and magnetic fields of the colorful region are observed in the continuum while on the lattice the electric field of the colorful region is only observed. In the $Q=0.5$ configuration, by substituting the colorful region within the circle of radius $R$ around the point ($x_1,z_1$), the orientation of the electric field within the circle corresponding to the unicolor vortex becomes opposite orientation due to the insertion of a circular monopole line around the intersection point \cite{Nejad:2016fcl}. Therefore, combining the electric field of circular monopole line of the $xy$-vortices and the magnetic field of the $zt$-vortices contributes $Q=-0.5$. Increasing the smoothing region $\Delta t$ of the colorful vortex, the monopole line which changes its color along the circle in a nontrivial way contributes itself with the value $Q=-1$ to the total topological charge. Therefore, increasing $\Delta t$ as well as increasing the radius $R$ of the colorful region through increasing the lattice size, the total topological charge converges to $Q=-1.5$, as shown in Fig.~\ref{fig:1}. 
 
\begin{figure}[h!] 
\centering
a)\includegraphics[width=0.4\columnwidth]{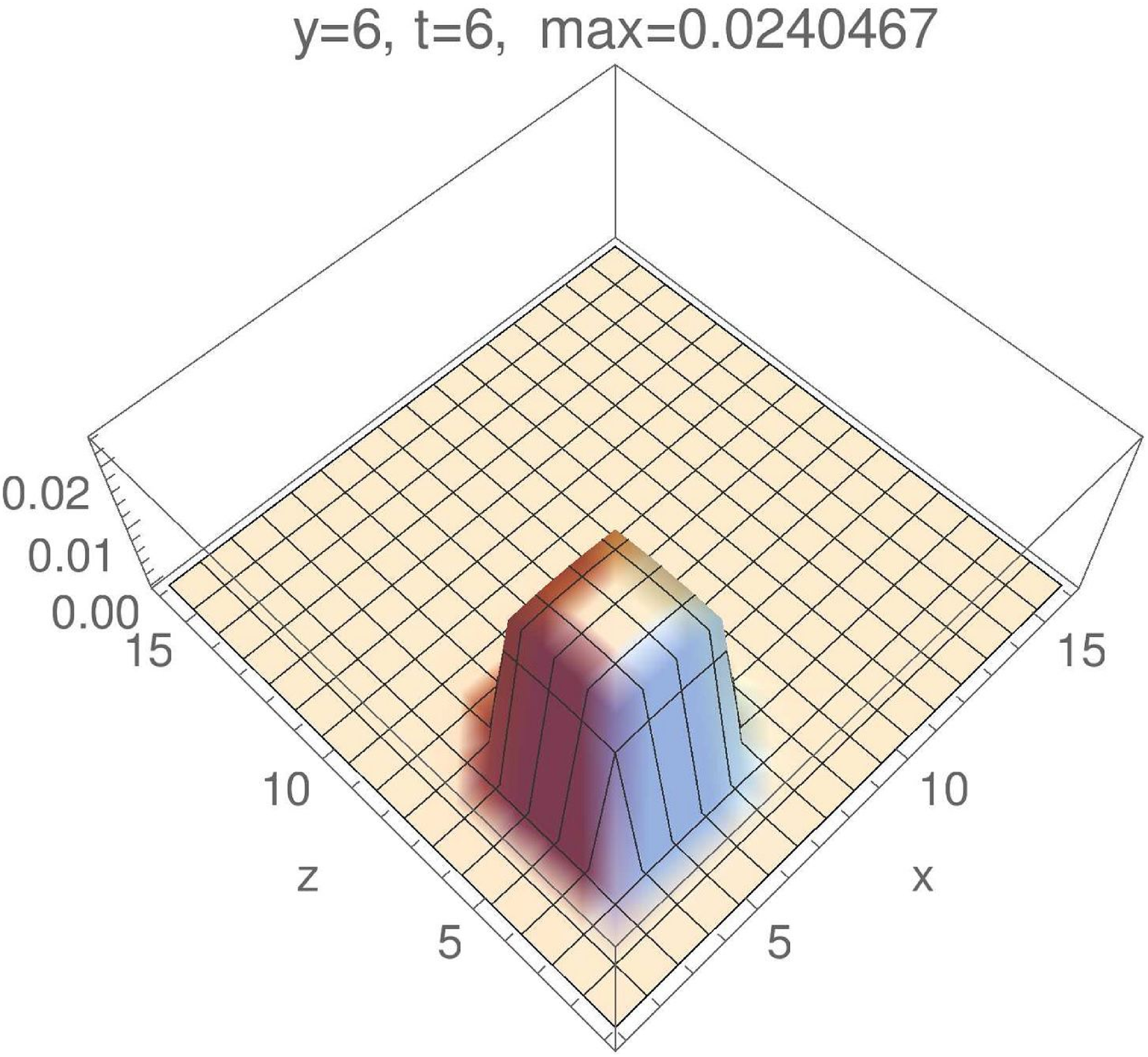}\hspace{1cm}
b)\includegraphics[width=0.4\columnwidth]{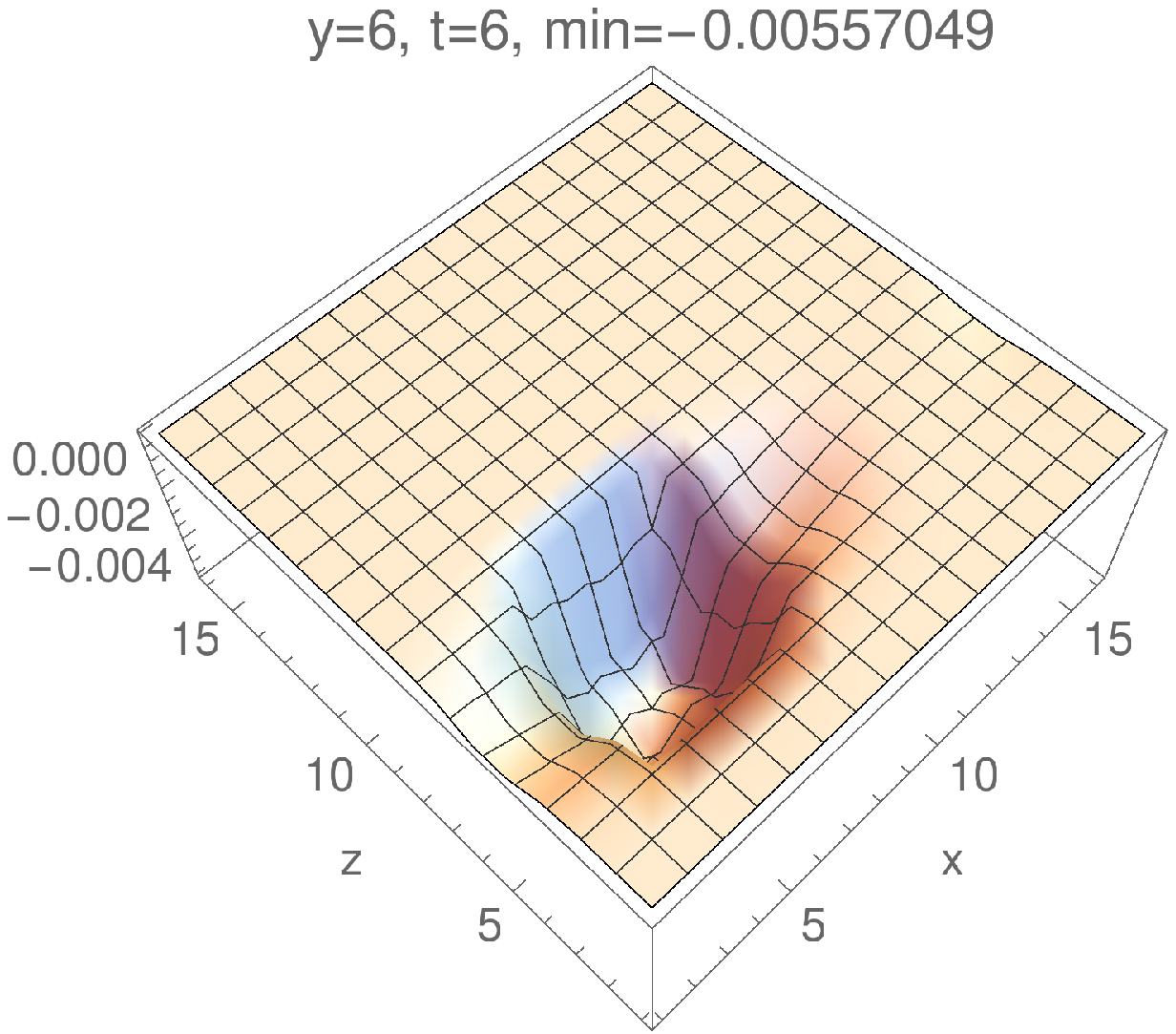}
\caption{(a) The topological charge density for the configuration in the xz-plane depicted in Fig.~\ref{fig:2}(a). Two antiparallel $xy$- and $zt$-vortex pairs are combined with vortex centers ($z_1=6,z_2=14$) and ($x_1=6,x_2=14$) at $t_\perp=y_\perp=6$ with thickness $d=2$ on a $16^4$-lattice. Intersecting vortex sheets related to the same color direction contribute to the topological charge. Therefore, only the intersection point around $(x_1,z_1)$, where the vortex sheets are related to $\sigma_3$, contributes to the topological charge. The intersection point around ($x_1=6,z_1=6$) gives rise to a lump of topological charge $Q=0.5$. (b) the configuration, depicted in Fig.~\ref{fig:2}(b), is the same as the left diagram but the unicolor region for the $xy$-vortices around the point ($x_1,z_1$) is substituted by a colorful region, negatively charged. The smoothing region is $\Delta t=t_2-t_1=9-3=R=6$. The intersection point around ($x_1=6,z_1=6$) gives rise to a lump of topological charge $Q=-1.5$.}
\label{fig:3}
\end{figure}

Topological charge densities of the $Q=0.5$ and $Q=-1.5$ configurations in the intersection plane show the details of the contribution to the topological charge. Figure~\ref{fig:3}a shows characteristic charge density for the $Q=0.5$ configuration with two intersecting antiparallel $xy$- and $zt$-vortex pairs at ($z_1=6,z_2=14$) and ($x_1=6,x_2=14$) at $t_\perp=y_\perp=6$ with thickness $d=2$ on a $16^4$-lattice. The intersection point around ($x_1=6,z_1=6$) gives rise
to a lump of topological charge $Q=0.5$. Figure~\ref{fig:3}b shows characteristic charge density for the $Q=-1.5$ configuration. The parameters of the configuration are the same as those of the Fig.~\ref{fig:3}(a). The center of the colorful region with radius $R=\Delta t=6$ in the $xy$ plane is located at $x_0=x_1=6,\;y_0=y_\perp=6$. The intersection point around ($x_1=6,z_1=6$) gives rise
to a lump of topological charge $Q=-1.5$.  

In the next section, we investigate the fundamental and adjoint representations of the overlap and asqtad staggered Dirac operator for these configurations with fractional topological charges.

\section{influence of fractional topological charges on Dirac modes}\label{Sect2}

In the previous section, we defined two configurations with the fractional topological charges which are combinations of
two antiparallel plane vortex pairs. For the $Q=0.5$ configuration, one unicolor intersection contributes to the topological charge while a colorful intersection contributes to the charge for the $Q=-1.5$ configuration. We calculate the lowest eigenmodes of the overlap and asqtad staggered Dirac operators in the fundamental and adjoint representations in order to study the influence of the fractional topological charges on fermions. According to the Atiyah-Singer index theorem for the overlap fermions in the fundamental representation, the integer topological charge is related to the index by $\mathrm{ind}D[A]=n_--n_+=Q$ where $n_-$ and $n_+$ denote the numbers of left- and right-handed zero modes 
\cite{Atiyah:1971rm,Brown:1977bj,Adams:2000rn}. In a generic topologically
nontrivial configuration, one may never find non zero values for both $n_-$ as well as
$n_+$, sometimes referred to as absence of fine tuning. However, the difference between these numbers agrees with the index theorem and further zero modes which cancel each
other in the index are non-topological zero modes. Therefore, the overlap Dirac operator in the fundamental representation for a configuration with integer topological charge $Q\neq 0$ has $|Q|$ topological zero modes with chirality $-\mathrm{sign} (Q)$. 

\begin{figure}[h!] 
\centering
a)\includegraphics[width=0.48\columnwidth]{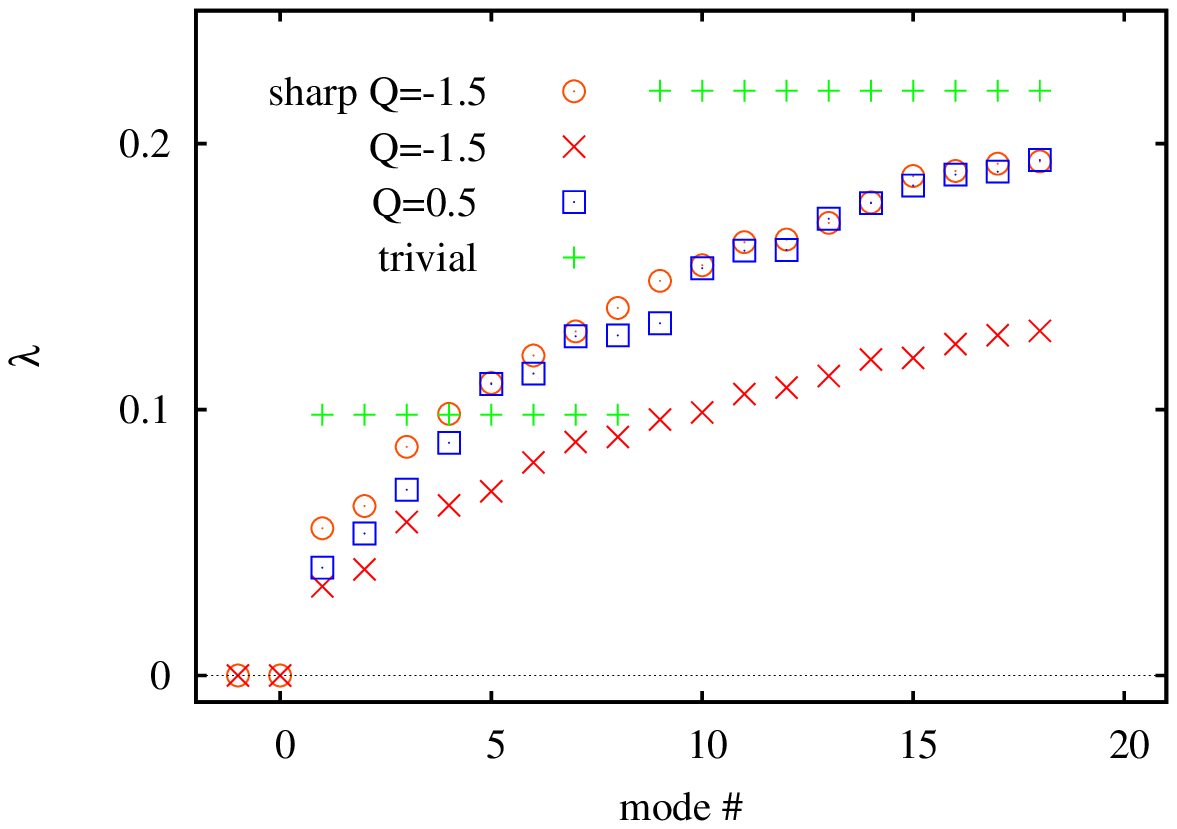}
b)\includegraphics[width=0.48\columnwidth]{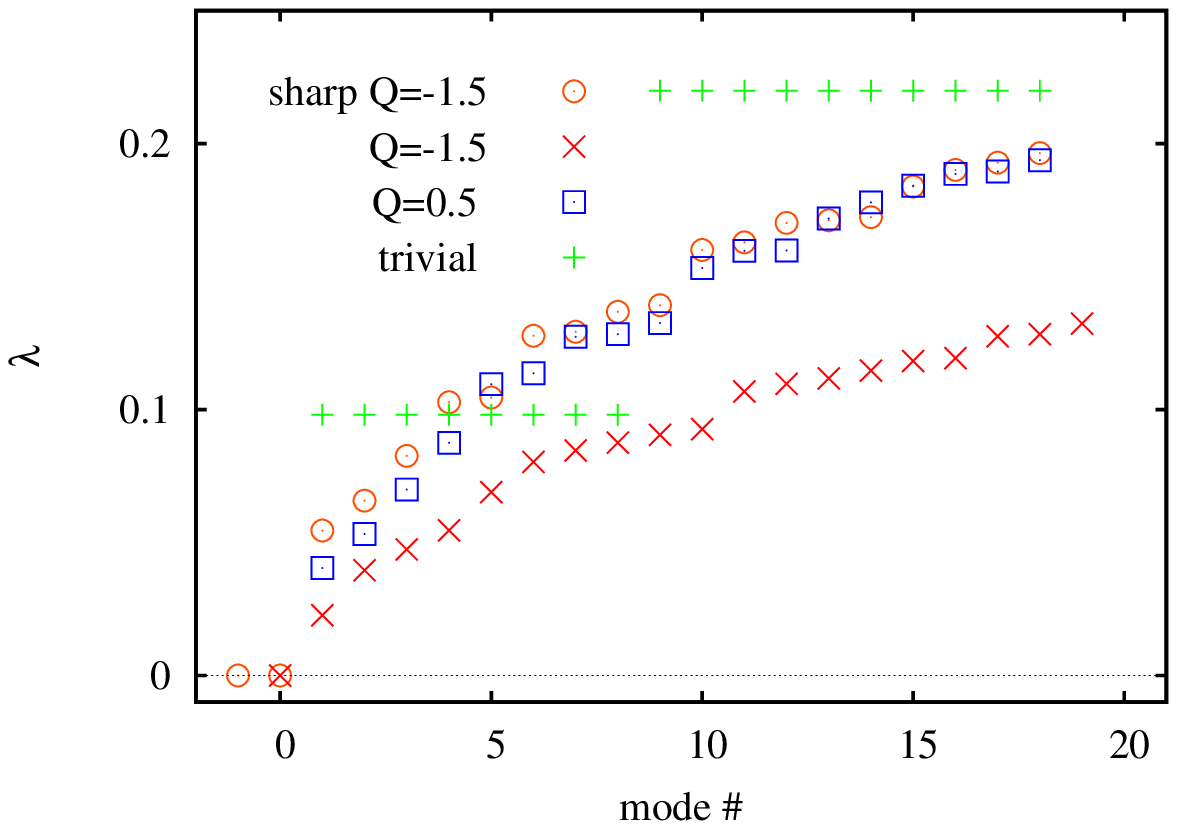}
\caption{ (a) The lowest eigenvalues of the overlap Dirac operator on a $16^4$ lattice in the fundamental representation for the vortex configurations with fractional topological charges schematically displayed in Fig.~\ref{fig:2} with $Q=0.5$ [Fig.~\ref{fig:2}(a)] and $Q=-1.5$ [Fig.~\ref{fig:2}(b)], using periodic bcs in spatial directions and antiperiodic bcs in temporal direction for fermionic fields. (b) the same as (a) but for periodic bcs in temporal direction. There are two zero modes for fundamental overlap operator on the sharp $Q=-1.5$ configuration which persist through changing the bcs, but just one of them remains zero after increasing the smoothing region $\Delta t$ of the colorful vortex. In the fundamental representation, we also get some low-lying eigenmodes for both configurations with smaller eigenvalues than the ones of the lowest eigenvectors for the trivial gauge field. These low lying modes could not be removed by changing the bcs.}\label{fig:4}
\end{figure}

\begin{figure}[h!] 
\centering 
a)\includegraphics[width=0.34\columnwidth]{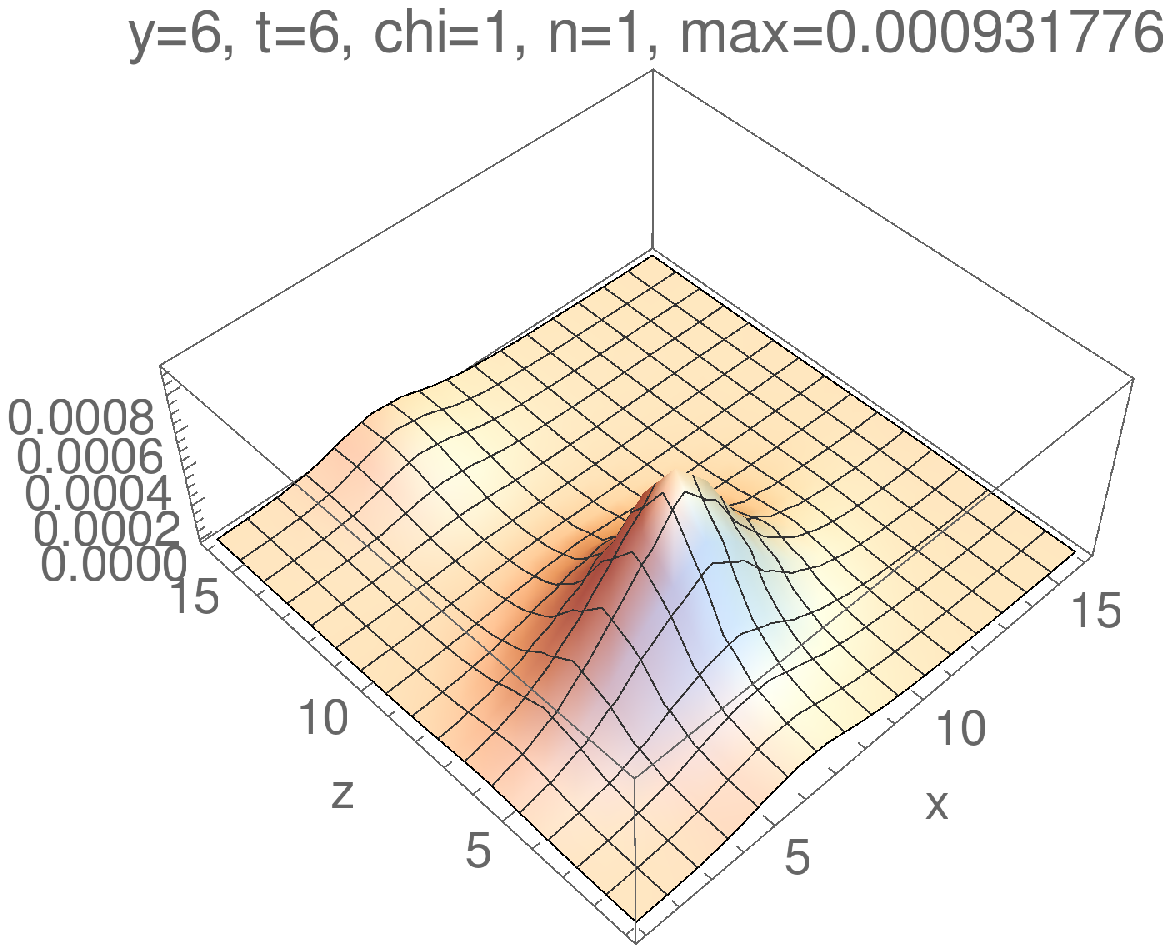}
b)\includegraphics[width=0.34\columnwidth]{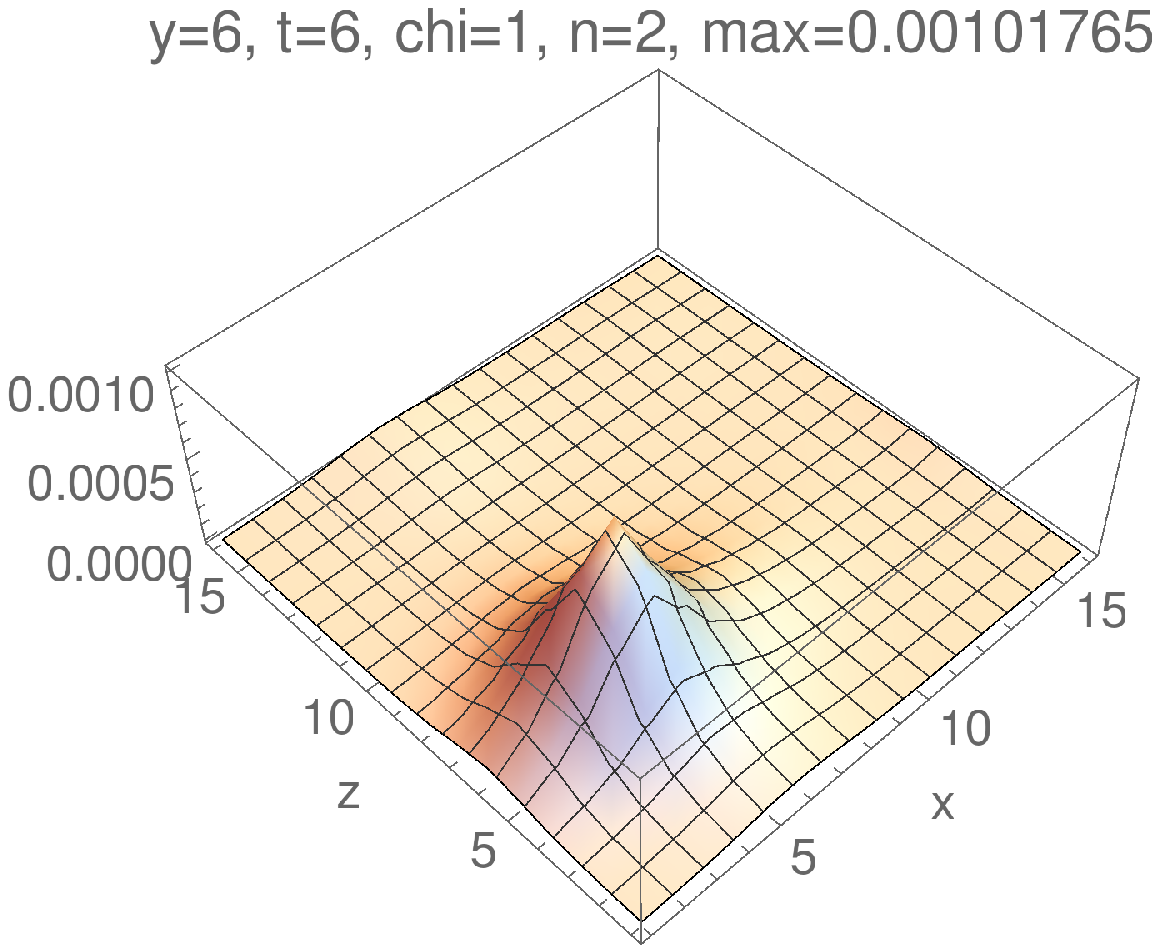}
c)\includegraphics[width=0.34\columnwidth]{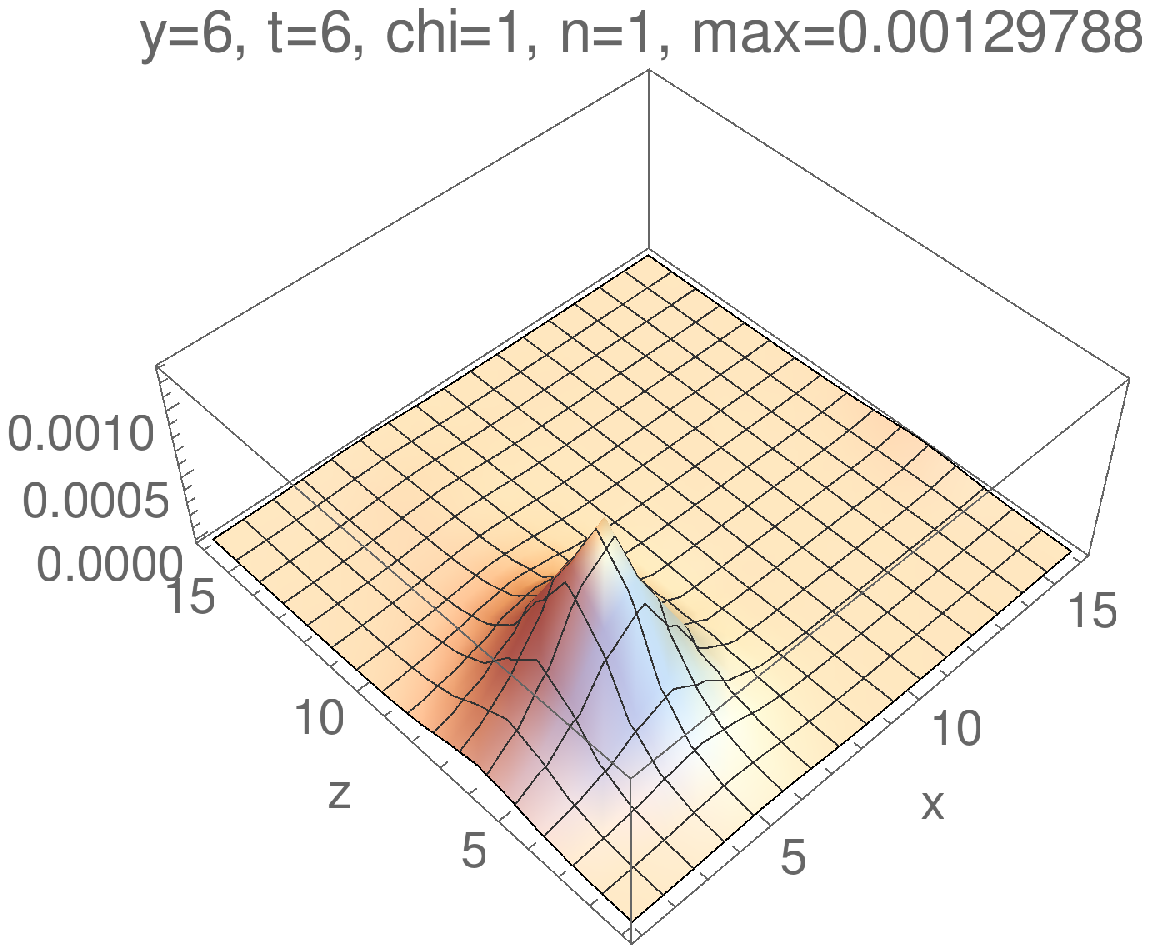}
d)\includegraphics[width=0.38\columnwidth]{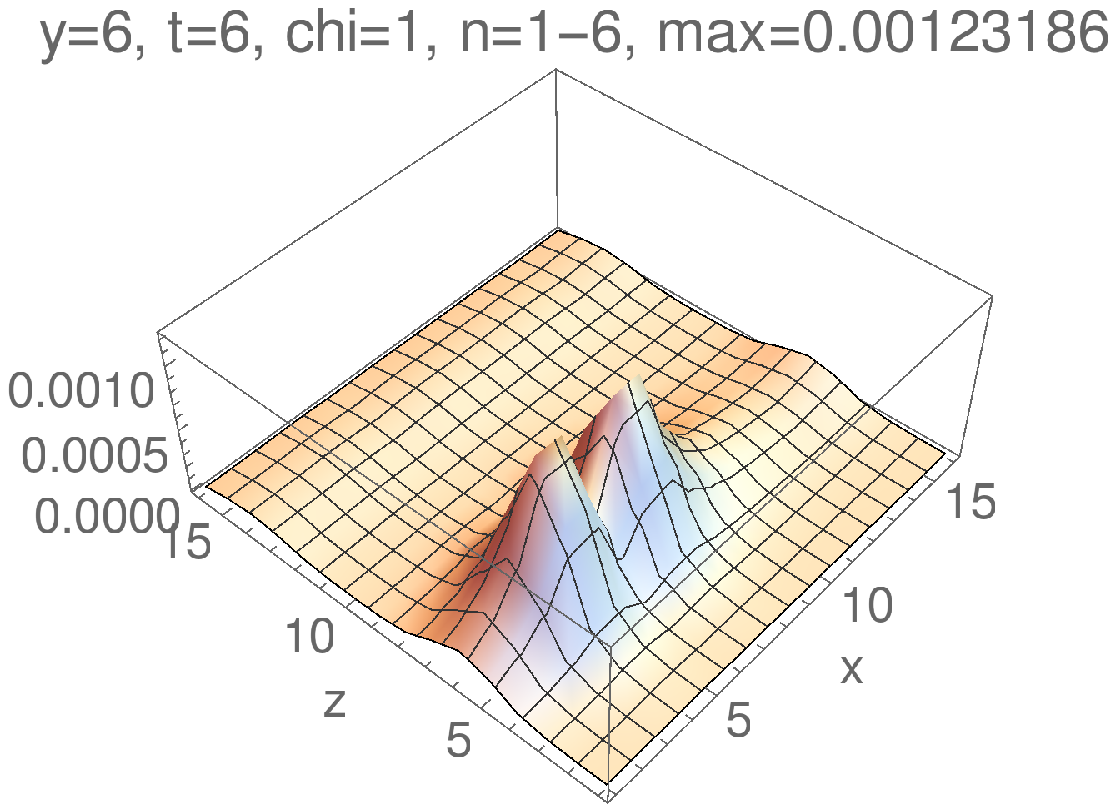}
\caption{ The chiral densities of a,b) two fundamental zero modes for antiperiodic bcs c) the zero mode for periodic bcs d) six adjoint zero modes for the $Q=-1.5$ configuration in the xz-plane on a $16^4$ lattice. Therefore there is one exact topological zero mode for fundamental overlap operator on the $Q=-1.5$ configuration. The plot titles indicate the plane positions, the chirality (chi=$0,\pm1$), the number $n$ of plotted mode and the maximal density in the plotted area, "max=...".}
\label{fig:5}
\end{figure}

In Figs.~\ref{fig:4}(a), and ~\ref{fig:4}(b), we show the lowest eigenvalues of the overlap Dirac operator in the fundamental representation for the two configurations compared to the eigenvalues of the free overlap Dirac operator for antiperiodic and periodic bcs on a $(16)^4$-lattice. The parameters of the configurations are the same as those in Fig.~\ref{fig:3}. As shown in Fig.~\ref{fig:4}(a), using periodic bcs in spatial directions and antiperiodic bcs in temporal directions for fermionic fields, we find two fundamental zero modes of positive chirality for the $Q=-1.5$ configuration and no fundamental zero-mode for the $Q=0.5$ configuration. Hence, the colorful intersection with fractional topological charge, which has a monopole loop, attracts two zero modes.  
Two fundamental zero modes in the background of the sharp $Q=-1.5$ configuration persist regardless of the bcs, but as shown in Fig. \ref{fig:4}(b) just one
of them remains zero in the background of the smooth $Q=-1.5$ configuration using periodic bcs in spatial and temporal directions. Therefore, there is one exact topological zero mode for fundamental overlap
operator on the $Q=-1.5$ configuration. In Figs.~\ref{fig:5}(a), ~\ref{fig:5}(b), and ~\ref{fig:5}(c), the chiral densities of zero modes for antiperiodic and periodic bcs are depicted in the xz-plane at $(y=6, t=6)$ which are localized at the colorful region. For the sharp $Q=-1.5$ configuration which two zero modes persist, we have singularity in time direction ($\Delta t=1$) where vortices are localized in a single time slice, but after smoothing over several lattice slices i.e. increasing the smoothing region $\Delta t$ of the colorful vortex, just one fundamental zero mode persist. 

\begin{figure}[h!] 
\centering
\includegraphics[width=0.48\columnwidth]{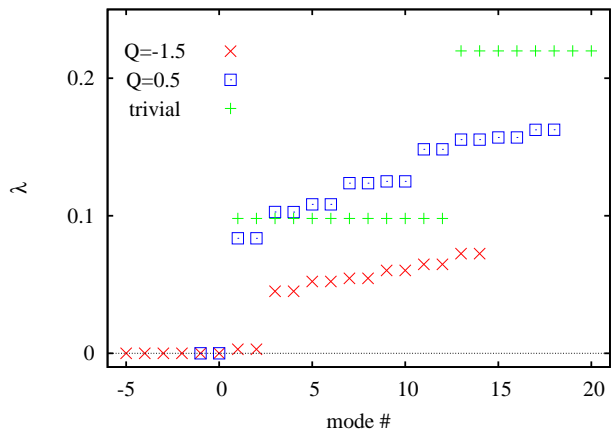}
\caption{ The lowest eigenvalues of the overlap Dirac operator on a $16^4$ lattice in the adjoint representation for the vortex configurations with fractional topological charges schematically displayed in Fig.~\ref{fig:2}. We find six adjoint zero modes of positive chirality for the $Q=-1.5$ configuration and two adjoint zero-modes of negative chirality for the $Q=0.5$ configuration which persist regardless of the bcs.}\label{fig:adjoint}
\end{figure}

In Fig.~\ref{fig:adjoint}, we show the lowest eigenvalues of the overlap Dirac operator in the adjoint representation for the noninteger $Q$ configurations compared to the eigenvalues of the free overlap Dirac operator on a $(16)^4$-lattice. The parameters of the configurations and fermionic fields are the same as those in Fig.~\ref{fig:4}(a). The overlap Dirac operator in the adjoint representation is sensitive to the topological charge contributions of $|Q|=0.5$ and attracts the zero modes which allow identification of fractional topological charge. According to the adjoint version of the Atiyah-Singer index theorem for the overlap fermions, the topological charge is related to the index by $\mathrm{ind}D[A]=n_--n_+=2NQ = 4Q$ where $N=2$ is the number of colors and the additional factor $2$ is due to the real representation of the fermion. As shown in Fig.~\ref{fig:adjoint}, we find six adjoint zero modes of positive chirality for the $Q=-1.5$ configuration which agrees with the index theorem. All adjoint zero modes are localized near to the colorful region. In Fig.~\ref{fig:5}(d), we show the combination of the chiral densities of six adjoint zero modes of the $Q=-1.5$ configuration. 

Next, we compare the number of left- and right-handed overlap zero modes with those of the asqtad staggered fermions. In Table~\ref{tab:result}, the fundamental and adjoint zero modes of the overlap and asqtad staggered Dirac operators are summarized. The results are taken more care regarding the continuum and infinite volume limits. There are two length scales in the configurations the thickness $d$ and the lattice extent $L$. The continuum limit would presumably correspond to taking $L \rightarrow \infty$ at a fixed $\zeta \equiv d/L$, and the infinite volume limit would
correspond to studying successively smaller values of $\zeta$ after taking the continuum limits. We used various lattice sizes up to $(28)^4$ and some fine thickness $d$ for studying the lattice spacing and finite volume effects to confirm the real numbers of zero modes. We will analyze in detail both effects on the low-lying modes of the overlap Dirac operator.  

\begin{table*}[!htbp]
\caption{\label{tab:result} Number of positive and negative overlap and asqtad staggered zero modes in
the fundamental and adjoint representations on the $Q=-1.5$ and $Q=0.5$ configurations. When the numbers of zero modes are different for periodic and antiperiodic bcs, they are shown in the table. We compared the results of various lattice sizes up to $(28)^4$ at a fixed $\zeta$ as well as some fine thickness d with smaller values of $\zeta$ to confirm the real numbers of zero modes.}
\begin{ruledtabular}
\begin{tabular}{lcccc}
&Ovl. fund. $(16)^4$ & Asq. stg. fund. $(28)^4$ & Ovl. adj. $(16)^4$ & Asq. stg. adj. $(28)^4$ \tabularnewline
\hline\hline
Sharp $Q=-1.5$  &2+0- & 4+0-& 6+0- & 12+0-\\
$Q=-1.5$  &apbc:2+0-pbc:1+0- & apbc:4+0-pbc:2+0-& 6+0- & 12+0-\\
$Q=0.5$  & 0+0- & 0+0-  & 0+2-  & 0+4-        \\
\end{tabular}
\end{ruledtabular}
\end{table*}

According to the index theorem for the (asqtad) staggered fermions, the integer topological charge is related to the index by $\mathrm{ind}D[A]=n_--n_+=2Q$ for the fundamental and $\mathrm{ind}D[A]=n_--n_+=8Q$ for the adjoint fermions. For the $Q=-1.5$ configuration, we find four and two asqtad staggered fundamental zero modes with positive chirality for antiperiodic and periodic bcs, respectively.  Therefore, there are two exact topological zero modes for the fundamental asqtad staggered
operator on the $Q=-1.5$ configuration. For the $Q=0.5$ configuration, there is no asqtad staggered fundamental zero mode. Also, we find twelve positive asqtad staggered adjoint zero modes for the $Q=-1.5$ configuration and four negative asqtad staggered adjoint zero modes for the $Q=0.5$ configuration which agree with the index theorem. For the $Q=-1.5$ configuration, the intersection and color structure contribute to the topological charge with $-0.5$ and $-1$. One overlap and two asqtad staggered fundamental zero modes for the $Q=-1.5$ configuration are related to the colorful contribution $-1$, as could be confirmed by Refs. \cite{Nejad:2015aia,Schweigler:2012ae}. It seems that $|Q|=0.5$ contribution of the colorful intersection does not attract any fundamental zero mode. 

As a result, it seems that fractional topological charge $|Q|=0.5$ does not attract fundamental zero modes while adjoint fermions on the other hand clearly identify the fractional topological charges according to the index theorem. 

In the continuum, the relation of the index theorem takes the form corresponding to the lattice overlap Dirac operator while the one corresponding to the lattice staggered Dirac operator has the additional factor $2$. In Table~\ref{tab:result}, the overlap and asqtad zero modes are the same after taking away staggered factor $2$. If one takes the continuum
limit of the background configurations, one expect to see the same spectrum for both the operators.

\begin{figure}[h!] 
\centering 
a)\includegraphics[width=0.48\columnwidth]{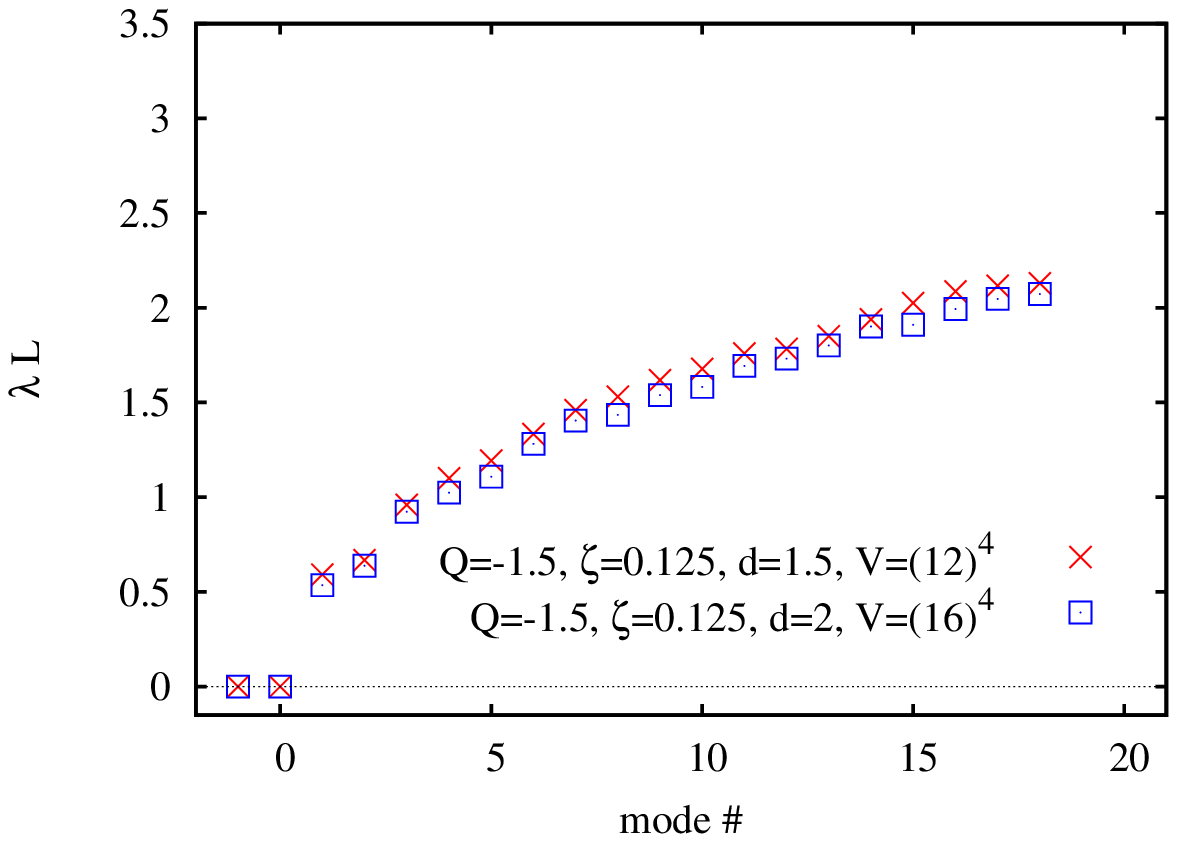}
b)\includegraphics[width=0.48\columnwidth]{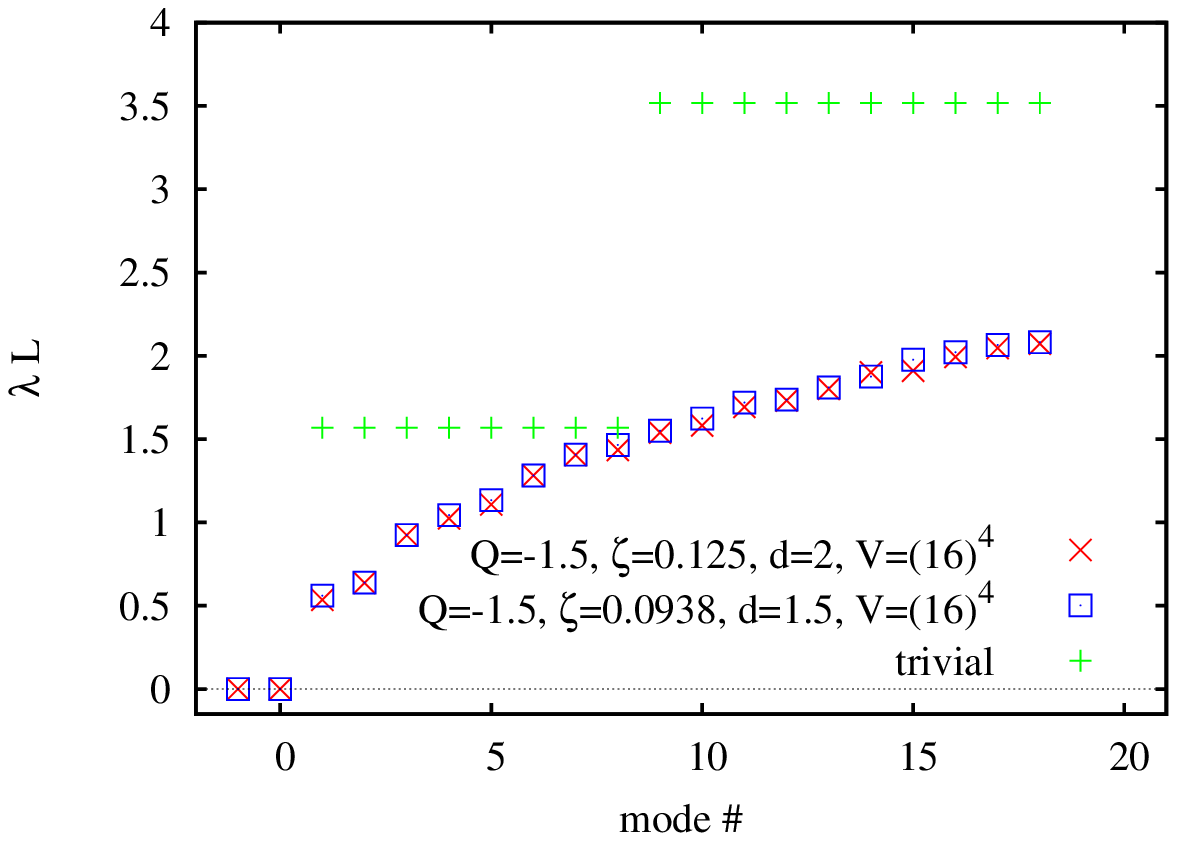}
\caption{ $\lambda L$ of low-lying modes of the overlap Dirac operator in the fundamental representation for some values of $\zeta \equiv d/L$
for the $Q=-1.5$ configuration (a) through increasing both length scales $L$ and $d$ at a fixed $\zeta$ for analyzing the lattice spacing effect (b) for analyzing the finite volume effect, the largest $L$ of a) is fixed while $d$ and therefore $\zeta$ are decreased. There
is a very slight difference between $L = 12$ and $L = 16$ data at fixed $\zeta$. Also, by varying the values of $\zeta$ at fixed $L$, the data stay without changes.}
\label{fig:zeta1}
\end{figure}

\begin{figure}[h!] 
\centering
a)\includegraphics[width=0.48\columnwidth]{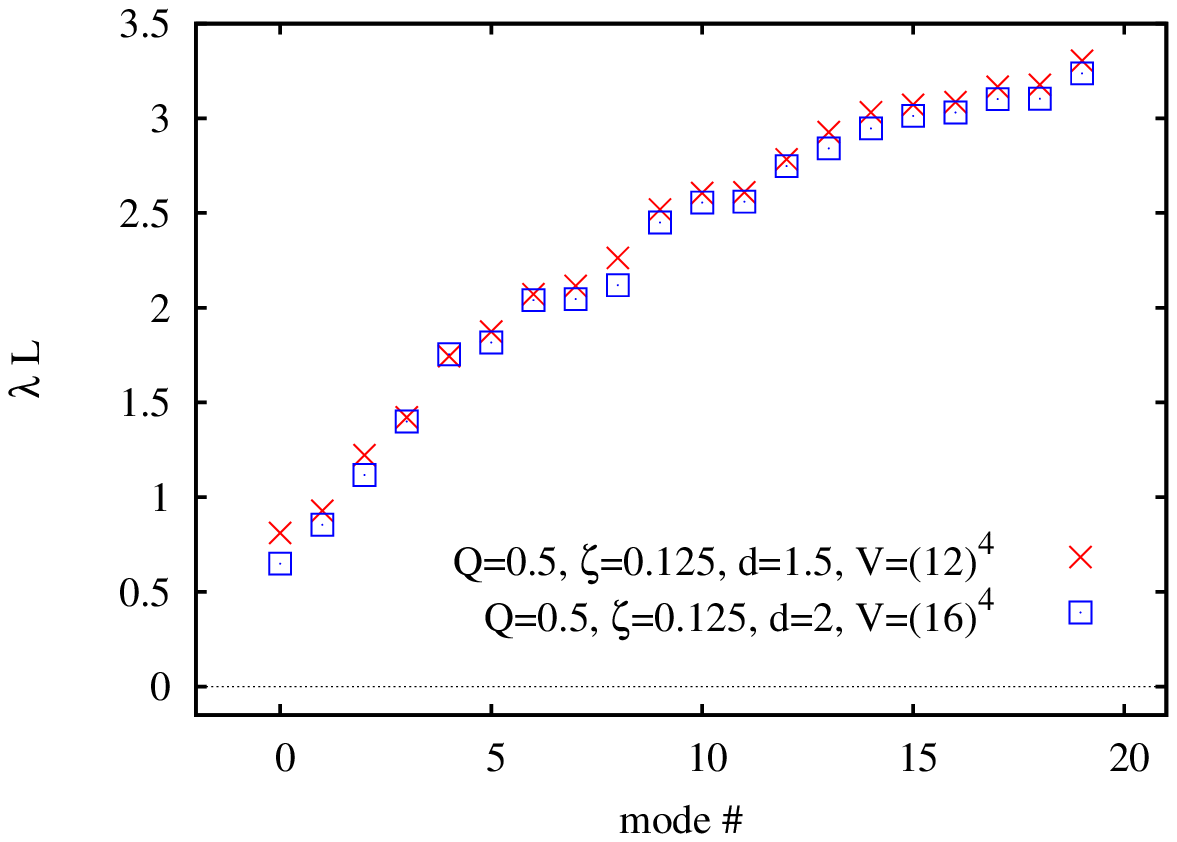}
b)\includegraphics[width=0.48\columnwidth]{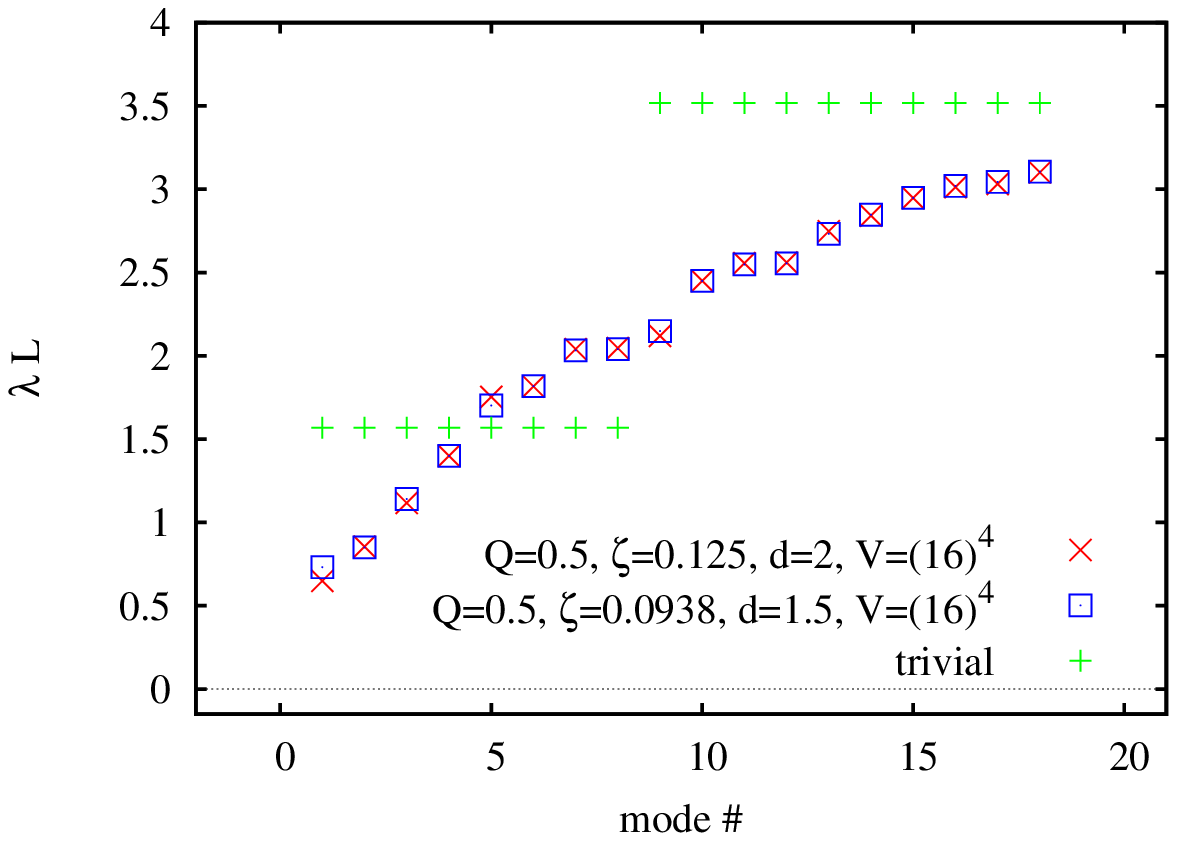}
\caption{ The same as Fig.~\ref{fig:zeta1} but for the $Q=0.5$ configuration. The behavior of $\lambda L$ of low-lying modes are the same as those of the $Q=-1.5$ configuration, plotted in Fig.~\ref{fig:zeta1}. }
\label{fig:zeta2}
\end{figure}

Further, we analyze the behavior of the low-lying (nonzero) modes of the overlap Dirac operator in the background of these configurations. In the fundamental representation, we get some lowest eigenmodes for both configurations with smaller eigenvalues than the ones of the lowest eigenvectors for the trivial gauge field. These low lying modes could not be removed by changing the bcs. For analyzing the behavior of these low-lying modes, the results are taken more care regarding the continuum and infinite volume limits. A systematic treatment of both
the lattice spacing and finite volume effects is necessary. If $\ell_{ph}$ is the length of the box in some physical
unit, then the lattice spacing is $\ell_{ph}/L$. The $\lambda$’s are the eigenvalues
in lattice units; so what is physical is $\lambda L= \lambda_{ph} \ell_{ph}$, where $\lambda_{ph}$ is the eigenvalue in physical
units. In Figs.~\ref{fig:zeta1} and \ref{fig:zeta2}, $\lambda L$ of low-lying modes for some values of $\zeta \equiv d/L$ are studied, where the lattice volume $V$ is defined to be $V=(L)^4$.  In Fig.~\ref{fig:zeta1}(a), we show the behavior of $\lambda L$ of the low-lying modes for the $Q=-1.5$ configuration through increasing both length scales $L$ and $d$ at a fixed $\zeta$ for analyzing lattice spacing effect. There
is a very slight difference between $L = 12$ and $L = 16$ data and therefore no lattice spacing effect on the low lying eigenvalues.

In Fig.~\ref{fig:zeta1}(b), for analyzing finite volume effect, the largest $L$ of Fig.~\ref{fig:zeta1}(a) is fixed while $d$ and therefore $\zeta$ are decreased. As shown, the values of the eigenvalues stay without changes. Therefore, there is no finite volume effect (the effect of $\zeta$) on the low lying eigenvalues.

Figure ~\ref{fig:zeta2} is similar to Fig.~\ref{fig:zeta1} but for the $Q=0.5$ configuration. The lattice spacing and finite volume effects on the fermions in the background of the $Q=0.5$ configuration are the same as those of the $Q=-1.5$ configuration. Therefore, there
is no lattice spacing and finite volume effects on the low lying eigenvalues. 

By the Banks-Casher formula, the chiral condensate, an order parameter for chiral
symmetry breaking, in a $4D$ box of volume $V$, taking the infinite volume limit, is proportional to density of near-zero modes of the Dirac operator as
\begin{equation}
\label{Banks}
\left\langle \bar{\psi}\psi \right\rangle=\frac{\pi \rho(0)}{V}.
\end{equation}
What causes the condensate is the inverse volume scaling of the gap between the
microscopic eigenvalues of the Dirac operator \cite{DeGrand:2009et,Akemann:1997wi}. To achieve SCSB, the eigenvalues of the Dirac operator must
accumulate, as the volume $V$ is taken to infinity, sufficiently fast near $\lambda=0$. A simple scaling argument
means that the accumulation must be such that the average level spacing $\Delta \lambda$ (gap) among the eigenvalues
must be roughly constant near $\lambda=0$, and inversely proportional to the volume $V$ \cite{Akemann:1997wi}.

\begin{figure}[h!] 
\centering
a)\includegraphics[width=0.48\columnwidth]{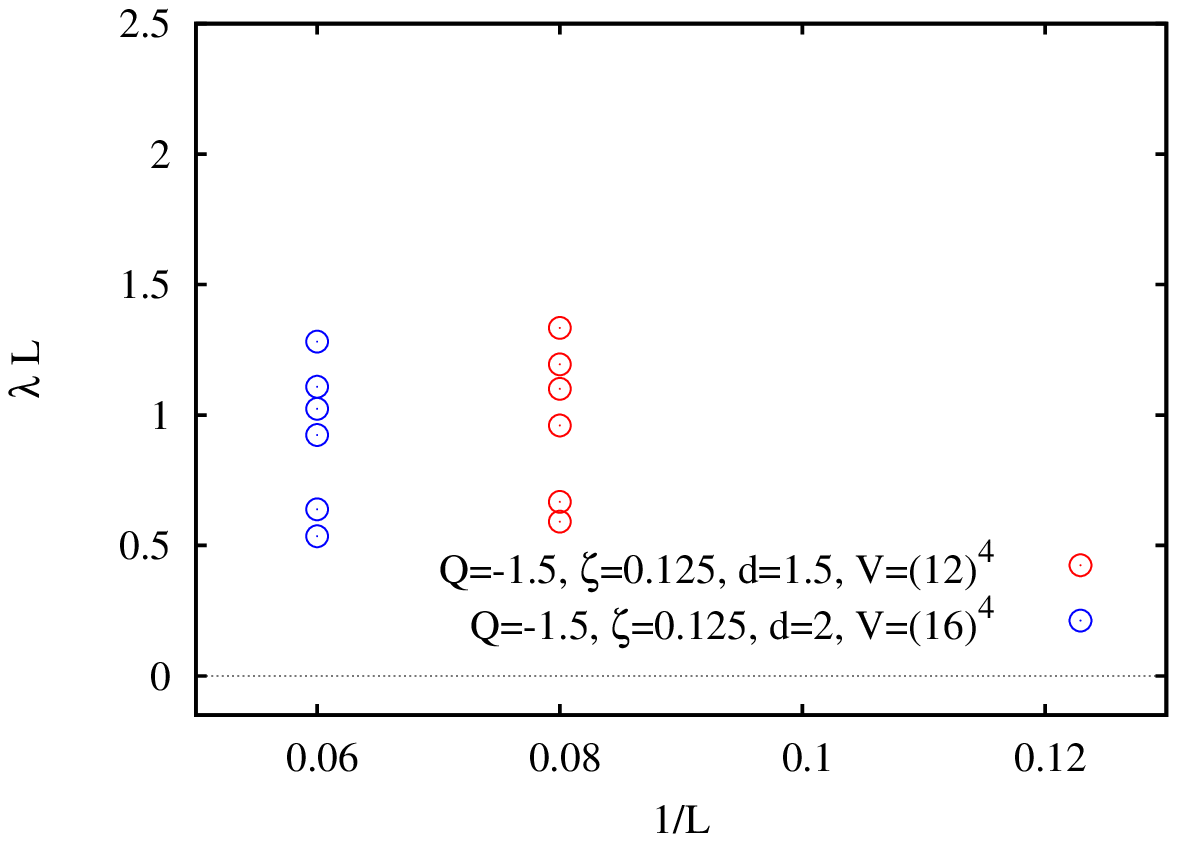}
b)\includegraphics[width=0.48\columnwidth]{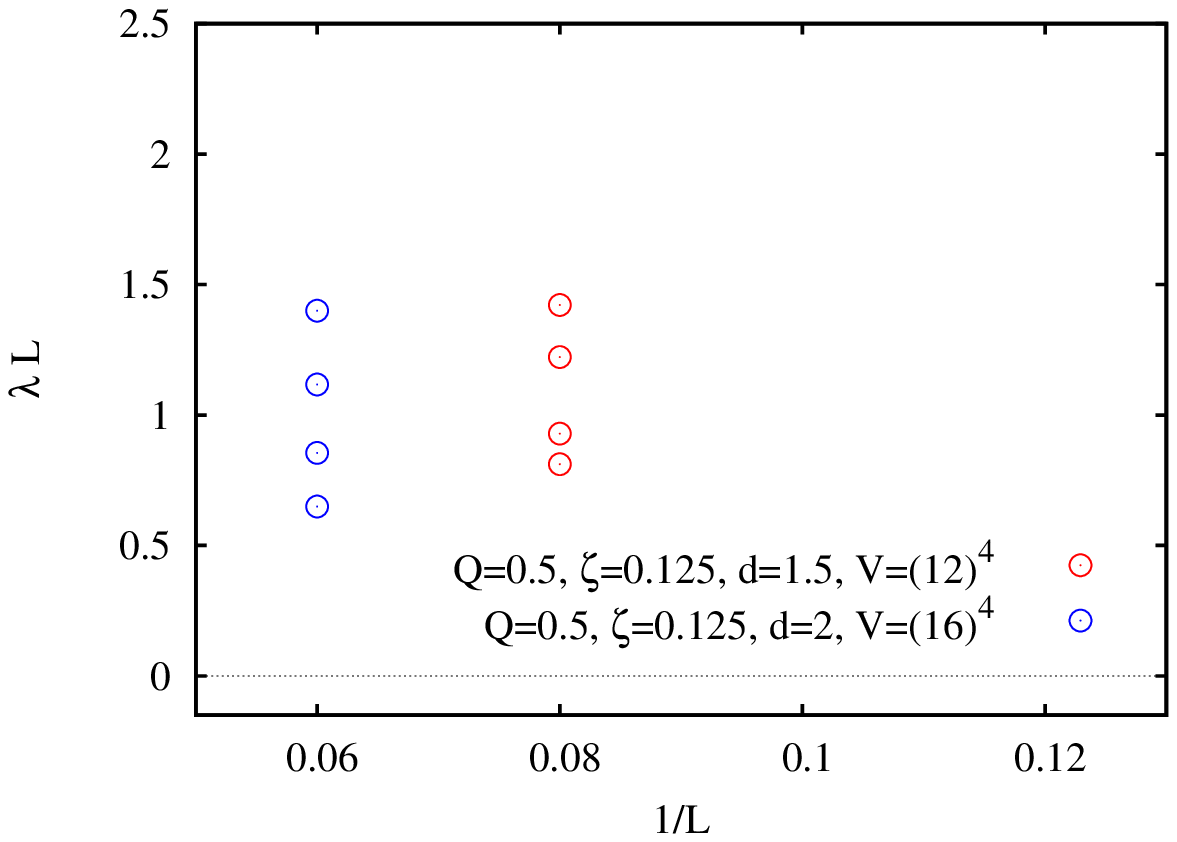}
\caption{ $\lambda L$ of low-lying (nonzero) modes of the overlap Dirac operator in the fundamental representation versus $1/L$ as the inverse volume scaling of the eigenvalues a) for the $Q=-1.5$ configuration b) for the $Q=0.5$ configuration. For both configurations, there
is no, or a very slight, difference between the $L = 12$ and $L = 16$ data. In fact, by increasing $L$, the gap between some eigenvalues becomes a little more. It seems that the lowest modes for the noninteger $Q$ configurations could not be identified as the near-zero modes. }
\label{fig:scaling}
\end{figure}

On the one hand, in Fig.~\ref{fig:scaling}, we show the behavior of $\lambda L$ of lowest modes of the overlap Dirac operator in the background of the noninteger $Q$ configurations versus $1/L$. As shown, there
is no or a very slight difference between $L = 12$ and $L = 16$ data. In fact, by increasing $L$, the gap between some eigenvalues becomes a little more. 

On the other hand, an inverse volume
scaling means $\lambda L \propto 1/\ell_{ph}^3$. But $\zeta \propto 1/\ell_{ph}$ at fixed physical distance between the vortices and therefore $\lambda L \propto \zeta^3$. As concluded in Figs.~\ref{fig:zeta1}(b) and \ref{fig:zeta2}(b), there is no dependence of $\lambda L$ on $\zeta$ at fixed $L$. This clearly rules out any inverse volume scaling of the microscopic eigenvalues.

In addition, although the lowest (nonzero) modes in the background of both configurations are smaller than the corresponding free field values, they are not extremely small. 

Therefore, it seems that the lowest (nonzero) modes for the noninteger $Q$ configurations which their eigenvalues are smaller than the free field values could not be identified as the near-zero modes and therefore no role for SCSB.  

\section{\boldmath Acknowledgments}
The author thanks the referee whose suggestions greatly improved this paper. 

\section{Conclusion}\label{Sect3}

Our aim is to understand the dynamical mechanism of nonperturbative phenomena of quark confinement and SCSB in QCD. Topological objects such as dyons with fractional topological charge could confine quarks and break chiral symmetry. Motivated by this idea, the fractional topological charge contributions from center vortices are investigated. We studied configurations with topological charges $Q=0.5$ and $Q=-1.5$, which are combinations of two antiparallel plane vortex pairs. For the $Q=0.5$ configuration, one unicolor intersection contributes to the topological charge while one colorful intersection carries the topological charge for the $Q=-1.5$ configuration. The colorful region in the configuration introduces a monopole line on its vortex surface surrounding the intersection point. 

We analyzed the low-lying modes of the Dirac operator in the background of the fractional topological charges through taking more care regarding the continuum and infinite volume limits. For the $Q=-1.5$ configuration, we find two overlap fundamental zero modes of positive chirality for the antiperiodic bcs. But using periodic bcs just one
of them remain zero. Therefore, there is one exact topological zero mode for fundamental overlap
operator on the $Q=-1.5$ configuration. Although, for the sharp $Q=-1.5$ configuration where we have singularity in time direction, two zero modes persist regardless of the bcs. However, after smoothing over several lattice slices, just one fundamental zero mode persist.
In addition, for the antiperiodic bcs, we find four asqtad staggered fundamental zero modes of positive chirality, while for the periodic bcs we get two zero modes. The adjoint zero modes allow identification of fractional topological charge. For this configuration, we get six overlap and twelve asqtad staggered adjoint zero modes of positive chirality which agree with the index theorem. For the $Q=0.5$ configuration, we get no overlap and no asqtad staggered fundamental zero modes. Also, for this configuration, two overlap and four asqtad staggered adjoint zero modes with negative chirality are found which again agree with the index theorem.  

As a result, it seems that fractional topological charge $|Q|=0.5$ does not attract fundamental zero modes while adjoint fermions on the other hand clearly identify this fractional topological charge  according to the index theorem.

There is no lattice spacing and finite volume effects on the low lying eigenvalues in the background of the vortex configurations with fractional topological charges. This clearly rules out any inverse volume scaling of the microscopic eigenvalues. In addition, although some lowest (nonzero) modes in the background of both configurations which are smaller than the corresponding free field values could not be removed by changing the bcs, they are not extremely small. Therefore, it seems that the lowest (nonzero) modes for the noninteger $Q$ configurations could not be identified as the near-zero modes and, therefore, the eigenvalues smaller than the free field values do not have a role for SCSB.

\bibliographystyle{unsrt}
\bibliography{paper}

\end{document}